\documentclass[twoside]{article}
\usepackage{qic,epsfig}
\usepackage{amsmath}
\begin{document}
\setlength{\textheight}{8.0truein}    

\runninghead{ GENUINE TRIPARTITE ENTANGLEMENT SEMI-MONOTONE FOR
$(2\times 2\times n)-$DIMENSIONAL SYSTEMS}
            {Chang-shui Yu, He-shan Song and Ya-hong Wang}

\normalsize\textlineskip \thispagestyle{empty} \setcounter{page}{1}

\copyrightheading{0}{0}{2003}{000--000}

\vspace*{0.88truein}

\alphfootnote

\fpage{1}

\centerline{\bf
GENUINE TRIPARTITE ENTANGLEMENT SEMI-MONOTONE }
\vspace*{0.035truein} \centerline{\bf FOR $(2\times 2\times
n)-$DIMENSIONAL SYSTEMS} \vspace*{0.37truein}
\centerline{\footnotesize
Chang-shui Yu$^1$, He-shan
Song$^1$\footnote{quaninformation@sina.com} and Ya-hong
Wang$^{1,2}$} \vspace*{0.015truein} \centerline{\footnotesize\it
$^1$Department of Physics, Dalian University of Technology, Dalian,
116024, P. R. China }
 \centerline{\footnotesize\it $^2$School of information science and engineering, Dalian Institute of
Light Industry, Dalian 116034, P. R. China }\baselineskip=10pt
\vspace*{10pt} \vspace*{0.225truein} \publisher{(received
date)}{(revised date)}

\vspace*{0.21truein}

\abstracts{
In this paper, we present a new approach to study genuine tripartite
entanglement existing in $(2\times 2\times n)-$dimensional quantum
pure states. By utilizing the approach, we introduce a particular
quantity to measure genuine tripartite entanglement. The quantity is
shown to be an entanglement monotone in 2-dimensional subsystems
(semi-monotone) and reaches zero for separable states and $(2\times
2\times 2)-$dimensional $W$ states, hence is a good criterion to
characterize genuine tripartite entanglement. Furthermore, the
formulation for pure states can be conveniently extended to the case
of mixed states by utilizing the kronecker product approximation
technique. As applications, we give the analytic approximation for
weakly mixed states, and study the genuine tripartite entanglement
of two given weakly mixed states. }{}{}

\vspace*{10pt}

 \vspace*{3pt} \communicate{to be filled by the Editorial}

\vspace*{1pt}\textlineskip    

\section{Introduction}

Entanglement is a valuable physical resource for many quantum
information processing, such as quantum computation [1], quantum
cryptography [2], quantum teleportation [3], quantum dense coding
[4] and so on. The understanding of entanglement is at the very
heart of quantum information theory. Recently, many efforts have
been made to characterize quantitatively the entanglement properties
of a quantum system[5-8], however, the good understanding is only
restricted in low-dimensional systems. The quantification of
entanglement for higher dimensional systems and multipartite quantum
systems remains to be an open question.

Since the remarkable concurrence was presented [5], it has been shown to be
a useful entanglement measure for the systems of qubits. Based on the
concurrence, V. Coffman et al [9] introduced the so called residual
entanglement for tripartite systems of qubits and shew that the residual
entanglement can be employed to measure genuine tripartite entanglement,
which opens the path to studying multipartite entanglement. However, unlike
the entanglement in low-dimensional systems, the entanglement in
high-dimensional or multipartite systems is much more complicated. E.g. D%
\"{u}r et al [10] have shown that three qubits can be entangled in
two inequivalent ways; Miyake [11,12] has shown that the
multipartite entanglement can be divided into more classes based on
the hyperdeterminant. These inequivalent entanglement classes tell
us that a single quantity can not effectively and thoroughly measure
entanglement of a high-dimensional or multipartite systems. However,
for some particular purpose, one can still characterize entanglement
by only a single quantity. For example: The most naturally, one can
measure entanglement of a certain class by only a single quantity
[11-14]; One can employ only a quantity to study the separable
property of a given quantum system [15,16]; One can also collect the
contributions of some different entanglements as a whole to study
the correlations between subsystems [17-20]; And so on.

In this paper, again a single quantity denoted by $\tau $ is presented in
terms of a new approach to characterize the genuine tripartite entanglement for $%
(2\times 2\times n)-$dimensional quantum systems. The distinct advantage of $%
\tau $ is that it can not only characterize the properties of
genuine tripartite entanglement existing in a given quantum pure
state and be conveniently extended to mixed states by kronecker
product approximation technique, but it is a entanglement
semi-monotone, i.e. it is an entanglement monotone considering the
two 2-dimensional subsystems and invariant under local unitary
transformations in the higher-dimensional subsystem. In this sense,
if the usual Positive Operation-Valued Measures (POVM's) on the
higher-dimensional subsystem is not considered, $\tau $ is even a
good entanglement measure. Furthermore, one will also find that the
initial residual entanglement introduced in Ref. [9] can be obtained
by our approach. In this sense, we also consider that $\tau $ is a
generalization of the initial residual entanglement. As
applications, we give the analytic approximation of $\tau $ for
weakly mixed tripartite quantum states (quasi pure states) and
consider the genuine tripartite entanglement of some quasi pure
states, which shows the sufficiency of our measure as a criterion to
test entanglement and the workability as an indicator of
entanglement in these cases. Note that even though there are other
results [12,21,22] for $(2\times 2\times n)-$dimensional quantum
systems, they are essentially different from ours. For example, Ref.
[21-22] studied the entanglement of assistance which is some kind of
bipartite entanglement in fact. Ref. [12] mainly focused on the
classification of multipartite entanglement and so far it seemed
difficult to obtain an operational entanglement evaluation for mixed
states. The paper is organized as follows. First, we give $\tau $
for pure states by a new approach and prove that $\tau $ is a
entanglement semi-monotone and can characterize genuine tripartite
entanglement; and then we extend it to mixed states and discuss the
genuine tripartite entanglement of some quasi pure states; the
conclusions are drawn in the end.

\section{The genuine tripartite entanglement semi-monotone for pure states}

\bigskip At first, let us introduce the concept of "tilde inner products".
The concept was presented by Wootters to introduce the remarkable
concurrence in Ref. [5]. Considering any two bipartite state vectors
of qubits $\left\vert x\right\rangle $ and $\left\vert
y\right\rangle $, the tilde inner product of $\left\vert
x\right\rangle $ and $\left\vert y\right\rangle $ is defined by
\begin{equation}
\left\langle x\right\vert \left. \tilde{y}\right\rangle =\left\langle
x\right\vert \sigma _{y}\otimes \sigma _{y}\left\vert y^{\ast }\right\rangle
,
\end{equation}%
where $\left\vert \tilde{y}\right\rangle =\sigma _{y}\otimes \sigma
_{y}\left\vert y^{\ast }\right\rangle $ with $\left\vert y^{\ast
}\right\rangle $ is the complex conjugate of $\left\vert y\right\rangle $
and $\sigma _{y}=\left(
\begin{array}{cc}
0 & -i \\
i & 0%
\end{array}%
\right) $. However, for convenience, whenever the tilde inner product is
mentioned, we refer to
\begin{equation}
\left( \left\langle x\right\vert \left. \tilde{y}\right\rangle \right)
^{\ast }=\left\langle x^{\ast }\right\vert \sigma _{y}\otimes \sigma
_{y}\left\vert y\right\rangle .
\end{equation}%
Now, let us focus on $\left( 2\times 2\times n\right) -$dimensional
tripartite quantum pure state $\left\vert \psi _{ABC}\right\rangle $ defined
in the Hilbert space $H_{1}\times H_{2}\times H_{3}$, which can be written
in the standard basis by
\begin{equation}
\left\vert \psi _{ABC}\right\rangle
=\sum_{i,j=0}^{1}\sum_{k=0}^{n-1}a_{ijk}\left\vert i\right\rangle
_{A}\left\vert j\right\rangle _{B}\left\vert k\right\rangle
_{C}=\sum_{k=0}^{n-1}\left\vert \varphi _{k}\right\rangle \left\vert
k\right\rangle ,
\end{equation}%
where $\left\vert \varphi _{k}\right\rangle
=\sum_{i,j=0}^{1}a_{ijk}\left\vert i\right\rangle _{A}\left\vert
j\right\rangle _{B}$ corresponds to $\left\vert k\right\rangle $ ($%
=\left\vert k\right\rangle _{C}$) of the party $C$. For any a group of basis
$\left\{ \left\vert \phi _{l}\right\rangle \right\} ,l=0,1,\cdot \cdot \cdot
,n-1$, defined in $H_{3}$, one can always project $\left\vert \psi
_{ABC}\right\rangle $ onto them, and obtain correspondingly an unnormalized
bipartite pure state of qubits defined in $H_{1}\times H_{2}$. Without loss
of generality, here we choose $\left\{ \left\vert \phi _{l}\right\rangle
\right\} =\left\{ \left\vert k\right\rangle \right\} $, therefore one can
obtain a corresponding set of unnormalized bipartite pure states $\left\{
\left\vert \varphi _{k}\right\rangle \right\} $. Arranging these bipartite
states in terms of the order of $\left\vert k\right\rangle $, we can
construct a matrix $\Phi $ given by%
\begin{equation}
\Phi =[\left\vert \varphi _{0}\right\rangle ,\left\vert \varphi
_{1}\right\rangle ,\cdot \cdot \cdot ,\varphi _{n-1}],
\end{equation}%
where it is implied that $\left\vert \varphi _{k}\right\rangle $ have been
considered as column vectors. Hence, on the basis of the tilde inner
product, we can obtain a new matrix $\mathcal{M}$ given by
\begin{equation}
\mathcal{M}=\Phi ^{T}\sigma _{y}\otimes \sigma _{y}\Phi ,
\end{equation}%
where each element $\mathcal{M}_{ij}$ denotes the tilde inner product of two
bipartite pure states $\left\vert \varphi _{i}\right\rangle $ and $%
\left\vert \varphi _{j}\right\rangle $. Furthermore, one will find
that $\mathcal{M}$ includes important information: First, because
each column of $\Phi$ is not normalized, $\Phi$ includes the
information of probabilities with which one can obtain the column
vectors from $\left\vert \psi _{ABC}\right\rangle $; Second, due to
the tilde inner product, $\mathcal{M}$, in particular its diagonal
elements, includes the separability information of the columns of
$\Phi$. Since $\mathcal{M}$ is a matrix defined in $\left( n\times
n\right) -$dimensional Hilbert space, we can always consider
$\mathcal{M}$ as a $\left( n\times n\right) -$dimensional bipartite
quantum pure state of qudits, which is given in matrix form.
Therefore, it is natural to consider the entanglement measure of
such an abstract bipartite pure state $\footnote{If $\mathcal{M}$ is
considered as an unnormalized pure state separately, it will make no
sense to measure the entanglement because there will exist a
undetermined constant (the normalization constant). However,
$\mathcal{M}$ is not separated here, but closely related to the
normalized $\left\vert \psi _{ABC}\right\rangle $. That is to say,
although the constant is not determined for $\mathcal{M}$, it is
determined for $\left\vert \psi _{ABC}\right\rangle $ which is what
we care for. Furthermore, the normalization constant of
$\mathcal{M}$ also includes valuable information mentioned in the
text and has its real value. Therefore, $\mathcal{M}$ should not be
normalized. }$.

As we know, I-concurrence [23,8] $C_{I}(\left\vert \Psi
\right\rangle )$ is a good measure for bipartite quantum pure state
$\left\vert \Psi \right\rangle $ defined in arbitrary dimension,
which has been shown to be the length of the concurrence vector [7]
by W. K. Wootters [24]. We can describe them by an equation as
\begin{equation}
C_{I}(\left\vert \Psi \right\rangle )=\sqrt{2[\left\vert \left\langle \Psi
\right\vert \left. \Psi \right\rangle \right\vert ^{2}-tr(\rho _{r}^{2})]}=%
\sqrt{\sum_{\alpha ,\beta =1}^{n(n-1)/2}\left\vert C_{\alpha \beta
}\right\vert ^{2}},
\end{equation}%
where $\rho _{r}$ denotes the reduced density matrix by tracing over one of
the two systems, $C_{\alpha \beta }=\left\langle \Psi ^{\ast }\right\vert
s_{\alpha }\otimes s_{\beta }\left\vert \Psi \right\rangle $ with $s_{\alpha
}$ and $s_{\beta }$ the generators of $SO(n)$. Applying the measure to the
state $\mathcal{M}$, we can obtain
\begin{equation}
C_{I}(\mathcal{M})=\sqrt{2\left\{ \left[tr\left(\mathcal{M}%
\mathcal{M}^{\dagger}\right)\right]^2-tr%
\left[ \left(\mathcal{M} \mathcal{M}^{\dagger} \right) ^{2}\right]\right\} }%
,
\end{equation}%
where $tr$ denotes trace operation.

As a consequence, we can summarize all above and obtain a following formal
expression.

\textbf{Theorem 1.} The genuine tripartite entanglement semi-
monotone $\tau $ for the given pure state $\left\vert \psi
_{ABC}\right\rangle $ can be obtained by
\begin{equation}
\tau =\tau \left( \left\vert \psi _{ABC}\right\rangle \right) =\left[ C_{I}(%
\mathcal{M})\right] ^{\frac{1}{2}},
\end{equation}%
where the corresponding parameters have been given above.

What's more, consider the expression of the second "$=$" in eq. (6),
we can
expand $\tau $ by%
\begin{eqnarray}
\tau &=&\left[ \sum_{i\neq l,j\neq k}\right. \left\vert \left\langle
\psi _{ABC}^{\ast }\right\vert \sigma _{y}\otimes \sigma _{y}\otimes
e_{ij}\left\vert \psi _{ABC}\right\rangle \right. \times
\left\langle \psi _{ABC}^{\ast }\right\vert \sigma _{y}\otimes
\sigma _{y}\otimes e_{lk}\left\vert \psi _{ABC}\right\rangle \notag \\
&&-\left\langle \psi _{ABC}^{\ast }\right\vert \sigma _{y}\otimes
\sigma _{y}\otimes e_{ik}\left\vert \psi _{ABC}\right\rangle \times
\left. \left. \left\langle \psi _{ABC}^{\ast }\right\vert \sigma
_{y}\otimes \sigma _{y}\otimes e_{lj}\left\vert \psi
_{ABC}\right\rangle \right\vert ^{2}\right] ^{\frac{1}{4}},
\end{eqnarray}%
where $e_{ij}=\left\vert i\right\rangle \left\langle j\right\vert $ with $%
\left\vert i\right\rangle $ denoting the standard basis of party $C$.

\textbf{Proof.} First of all, one can easily justify that:

$(\ast )$ $\tau $ given in above procedure can be reduced to the $\left(
2\times 2\times 2\right) $-dimensional case given in Ref. [9]. Noting that a
constant difference is neglectable.

Next, we will prove the theorem by two steps. At first, we will prove that $%
\tau $ is an entanglement semi-monotone, and then we will show that
$\tau $
can characterize genuine tripartite entanglement for $%
\left( 2\times 2\times n\right) -$dimensional pure state $\left\vert
\psi _{ABC}\right\rangle .$

\textit{Entanglement semi-monotone. }We first note that $\tau$ is
invariant under permutations of the two parties $A$ and $B$ of
$\left\vert \psi _{ABC}\right\rangle $ defined in 2-dimensional
Hilbert space respectively, hence we employ the method given in Ref.
[10] to prove that $\tau $ is non-increasing under local operations
assisted with classical communication (LOCC) in party $A$ only. Due
to the same reason mentioned in Ref. [10], we also consider a
sequence of two-outcome POVM's. Let $A_{1}$ and $A_{2}$ be the two
POVM elements such
that A$_{1}^{\dag }A_{1}+A_{2}^{\dag }A_{2}=\mathbf{1}_{2}$, with $\mathbf{1}%
_{\delta }$ denoting $\delta $-dimensional identity matrix, then $%
A_{i}=U_{i}D_{i}V$, where $U_{i}$ and $V$ are unitary matrices and $D_{i}$
are diagonal matrices with entries $(a,b)$ and $[\sqrt{1-a^{2}},\sqrt{1-b^{2}%
}]$, respectively. For some tripartite initial state $\left\vert \Psi
\right\rangle $, let $\left\vert \Theta _{i}\right\rangle =(A_{i}\otimes
\mathbf{1}_{2}\otimes \mathbf{1}_{n})\left\vert \Psi \right\rangle $ be the
unnormalized states obtained after the POVM operations. The corresponding
normalized states can be given by $\left\vert \Psi _{i}^{\prime
}\right\rangle =\left\vert \Theta _{i}\right\rangle /\sqrt{p_{i}}$, where $%
p_{i}=\left\langle \Theta _{i}\right\vert \left. \Theta _{i}\right\rangle $.
Then

\begin{equation}
\left\langle \tau \right\rangle =p_{1}\tau (\left\vert \Psi _{1}^{\prime
}\right\rangle )+p_{2}\tau (\left\vert \Psi _{2}^{\prime }\right\rangle ).
\end{equation}%
Considering the expression of a tripartite quantum state given in eq. (3), $%
\left\vert \Psi _{i}^{\prime }\right\rangle $ can also be rewritten by%
\begin{equation}
\left\vert \Psi _{i}^{\prime }\right\rangle =\sum_{k=0}^{n-1}\left( \frac{%
\left(A_{i}\otimes \mathbf{1}_{2}\right)\left\vert \varphi
_{k}\right\rangle }{\sqrt{p_{i}}}\right) \left\vert k\right\rangle .
\end{equation}%
Hence, after operation $A_{1}$, $\mathcal{M}^{\prime }$ corresponding to the
tilde inner products can be constructed by%
\begin{eqnarray}
\mathcal{M}_{jk}^{\prime } &=&\frac{1}{p_{1}}\left\langle \varphi
_{j}^{\ast }\right\vert \left[\left(
V^{T}D_{1}U_{1}^{T}\right)\otimes \mathbf{1_2} \right]\sigma
_{y}\otimes \sigma
_{y}\left[\left( U_{1}D_{1}V\right)\otimes \mathbf{1_2} \right] \left\vert \varphi _{k}\right\rangle  \notag \\
&=&\pm\frac{ab}{p_{1}}\left\langle \varphi _{j}^{\ast }\right\vert
\sigma
_{y}\otimes \sigma _{y}\left\vert \varphi _{k}\right\rangle =\pm \frac{ab}{%
p_{1}}\mathcal{M}_{jk}.
\end{eqnarray}%
Namely,
\begin{equation}
\mathcal{M}^{\prime }=\pm \frac{ab}{p_{1}}\mathcal{M}.
\end{equation}%
Analogously, $\mathcal{M}^{\prime \prime }$ corresponding to $A_{2}$ can be
given by%
\begin{equation}
\mathcal{M}^{\prime \prime }=\pm \frac{\sqrt{\left( 1-a^{2}\right) \left(
1-b^{2}\right) }}{p_{2}}\mathcal{M}.
\end{equation}%
On the basis of eq. (7) and eq. (8), one can obtain
\begin{equation}
\tau (\left\vert \Psi _{1}^{\prime }\right\rangle )=\frac{ab}{p_{1}}\tau
(\left\vert \Psi \right\rangle ),\tau (\left\vert \Psi _{2}^{\prime
}\right\rangle )=\frac{\sqrt{\left( 1-a^{2}\right) \left( 1-b^{2}\right) }}{%
p_{2}}\tau (\left\vert \Psi \right\rangle ).
\end{equation}%
Substituting eq. (15) into eq. (10), according to Ref. [10], one can
obtain that $\left\langle \tau \right\rangle \leq \tau (\left\vert
\Psi \right\rangle ).$ What's more, eq. (12) implies that there may
be an overall phase difference for $\mathcal{M}$ if a local unitary
transformation on party $A$ is considered. That is to say, $\tau
(\left\vert \Psi \right\rangle )$ will be invariant under such local
unitary transformations.

Now, let us focus on the third party $C$. Any a given $n\times n$ matrix $Q$
operated on party $C$ of $\left\vert \psi _{ABC}\right\rangle $ can be
described by $\left( \mathbf{1}_{2}\otimes \mathbf{1}_{2}\otimes Q\right)
\left\vert \psi _{ABC}\right\rangle $ denoted by $\left\vert \psi
_{ABC}^{\prime }\right\rangle $. Based on the tilde inner products, one can
always construct the corresponding matrix $\mathcal{\tilde{M}}$ following
above procedure. Consider the standard basis $\left\{ \left\vert
k\right\rangle \right\} $ of party $C$ in $H_{3}$, $\mathcal{\tilde{M}}$ can
be written by

\begin{equation}
\mathcal{\tilde{M}}_{ij}=\sum_{l,m=0}^{n-1}\left\langle \varphi _{l}^{\ast
}\right\vert \sigma _{y}\otimes \sigma _{y}\left\vert \varphi
_{m}\right\rangle Q_{il}Q_{jm},
\end{equation}%
where $\left\vert \varphi _{j}\right\rangle $ are defined the same to those
in eq. (3). If we operate $Q$ on $\mathcal{M}$ by $Q^{T}\mathcal{M}Q$,
considering the same basis $\left\{ \left\vert k\right\rangle \right\} $,
one can obtain
\begin{equation}
\left[ Q^{T}\mathcal{M}Q\right] _{lm}=\sum_{i,j=0}^{n-1}\left\langle \varphi
_{i}^{\ast }\right\vert \sigma _{y}\otimes \sigma _{y}\left\vert \varphi
_{j}\right\rangle Q_{il}Q_{jm},
\end{equation}%
where $\left[ \cdot \right] _{lm}$ denote the entries of the
corresponding matrix. Hence, from eq. (16) and eq. (17), one can get
\begin{equation}
\mathcal{\tilde{M}=}Q^{T}\mathcal{M}^{T}Q.
\end{equation}%
Since $\mathcal{M}$ can be regarded as a bipartite pure state in
matrix
form, we can assume that $\mathcal{M}$ is defined in the Hilbert space $%
H_{3}\otimes H_{3}^{\prime }$ and denotes an entangled state of
parties $C$ and $C^{\prime }$. According to eq. (18), we can draw a
conclusion that for the matrix $\mathcal{M}$ of the tilde inner
products, operating a transformation $Q$ on party $C$ of $\left\vert
\psi _{ABC}\right\rangle $ is
equivalent to operating $Q^{T}\otimes Q^{T}$ on the abstract state $\mathcal{%
M}^{T}$ which is defined in $H_{3}^{\prime }\otimes H_{3}$. In other words,
considering the local operations on party $C$ of $\left\vert \psi
_{ABC}\right\rangle $ is equivalent to considering the local operations on $%
\mathcal{M}^{T}$. If $Q$ is a unitary transformation, one can easily
find that the entanglement of $\mathcal{M}$ measured by eq. (7) is
invariant, i.e. $\tau $ is invariant under local unitary
transformation. If $Q$ is a usual POVM, one will not ensure that
$\tau $ is not always increasing. That is to say, $\tau$ is an
entanglement semi-monotone. This completes the first step.

\textit{Characterizing genuine tripartite entanglement. }Let us first show that $%
\tau =0$ for semiseparable pure states and low-local-rank $W$ states [11].
Considering the invariant permutation of $A$ and $B$, any semiseparable pure
state $\left\vert \psi _{ABC}\right\rangle $ can be given by
\begin{equation}
\left\vert \psi _{ABC}\right\rangle =\left\vert \varphi _{AB}\right\rangle
\otimes \left\vert \chi _{C}\right\rangle
\end{equation}%
or%
\begin{equation}
\left\vert \psi _{ABC}^{\prime }\right\rangle =\left\vert \chi
_{A}\right\rangle \otimes \left\vert \varphi _{BC}\right\rangle ,
\end{equation}%
where $\left\vert \chi _{i}\right\rangle $ denote the quantum pure
states for the $i$th single party and $\left\vert \varphi
_{pq}\right\rangle $ denote the bipartite pure states for the $p$th
and the $q$th parties. If projecting the state $\left\vert \psi
_{ABC}\right\rangle $ given by eq. (19) onto any a group of basis of
$H_{3}$ corresponding to party $C$, one can obtain that the
corresponding matrix $\mathcal{M}$ of the tilde inner
products has the entries given by%
\begin{equation}
\mathcal{M}_{ij}=\left\langle i\right. \left\vert \chi _{C}\right\rangle
\left\langle j\right. \left\vert \chi _{C}\right\rangle \left\langle \varphi
_{AB}^{\ast }\right\vert \sigma _{y}\otimes \sigma _{y}\left\vert \varphi
_{AB}\right\rangle .
\end{equation}%
According to eq. (7) and eq. (8), it is obvious that $\tau =0$ in
this case. If projecting the state $\left\vert \psi _{ABC}^{\prime
}\right\rangle $ given by eq. (20) onto any a group of basis
$\left\{ \left\vert \phi _{l}\right\rangle \right\} $ of $H_{3}$,
one can obtain the corresponding matrix
\begin{equation}
\mathcal{M}_{ij}^{\prime }=\left( \left\langle \chi _{A}^{\ast
}\right\vert \otimes \left\langle \kappa _{Bi}^{\ast }\right\vert
\right) \sigma _{y}\otimes \sigma _{y}\left( \left\vert \chi
_{A}\right\rangle \otimes \left\vert \kappa _{Bj}\right\rangle
\right) =0,
\end{equation}%
where $\left\vert \kappa _{Bj}\right\rangle =\left( \mathbf{1}_{2}\otimes
\left\langle \phi _{j}\right\vert \right) \left\vert \varphi
_{BC}\right\rangle $. Therefore, one can easily obtain that $\tau =0$ for $%
\left\vert \psi _{ABC}^{\prime }\right\rangle $. The $W$ states, the local
rank of which being $(2,2,2)$ is required, can always be reduced to a $%
(2\times 2\times 2)$-dimensional subspace of $H_{1}\otimes H_{2}\otimes
H_{3} $ and be considered as tripartite pure states of qubits. According to $%
(\ast )$, one can have $\tau =0$ for such $W$ states. It should be
noted that for the $W$ states with high local rank, we believe they
own genuine tripartite entanglement [25]. It is reasonable. As we
know, Ref. [11] has introduced the onionlike classification of
multipartite quantum states. The classification shows that the
quantum states in the outer class can always be converted
irreversibly into those in the inner class. Hence, we can say the
outer classes "include" the inner ones. $GHZ$ class with local rank
$(2,2,2)$ as the innermost class to characterize genuine tripartite
entanglement is hence "included" by outer class. In this sense, we
can safely say that $\tau \neq
0 $ for the $W$ states with high local rank. It is also in this sense that $%
\tau $ can be believed to be the generalization of the initial residual
entanglement.

Next, we will show that $\tau \neq 0$ for any quantum state with
genuine tripartite entanglement. According to the tensor treatment
of $\left\vert \psi _{ABC}\right\rangle $ [15], $\left\vert \psi
_{ABC}\right\rangle $ can be regarded as a tensor grid, whose units
can be considered to be tensor cubic. If there exist genuine
tripartite entanglements in $\left\vert \psi _{ABC}\right\rangle $,
there must exist at least such a tensor cubic of the grid as has
genuine tripartite entanglement. In other words, based on eq. (6),
there must exist some integers $\alpha ^{\ast }$ and $\beta ^{\ast
},$ such that $\left\vert C_{\alpha ^{\ast }\beta ^{\ast
}}\right\vert ^{2}\neq 0$. Hence, we can draw the conclusion that
$\tau =0$
means that there does not exist any genuine tripartite entanglement in $%
\left\vert \psi _{ABC}\right\rangle $. This completes the second
step.\square\
\section{Extension to mixed states}

\bigskip Consider $\tau \left( \left\vert \psi _{ABC}\right\rangle \right) $
of pure states, the corresponding quantity of mixed states $\rho $
is then given as the convex roof
\begin{equation}
\tau (\rho )=\inf \sum_{i}p_{i}\tau (\left\vert \gamma _{i}\right\rangle )
\end{equation}%
of all possible decompositions into pure states $\left\vert \gamma
_{i}\right\rangle $ with
\begin{equation}
\rho =\sum_{i}p_{i}\left\vert \gamma _{i}\right\rangle \left\langle \gamma
_{i}\right\vert ,p_{i}\geq 0.
\end{equation}%
$\tau (\rho )$ vanishes if and only if $\rho $ does not include any
genuine tripartite entanglement. According to the matrix notation
[7] of equation (24), one can obtain $\rho =\Gamma W\Gamma ^{\dagger
}$, where $W$ is a diagonal matrix with $W_{ii}=p_{i}$, the columns
of the matrix $\Gamma $ correspond to the vectors $\left\vert \gamma
_{i}\right\rangle $. Due to the eigenvalue decomposition: $\rho
=\Phi M\Phi ^{\dagger }$, where $M$ is a
diagonal matrix whose diagonal elements are the eigenvalues of $\rho $, and $%
\Phi $ is a unitary matrix whose columns are the eigenvectors of $\rho $,
one can obtain $\Gamma W^{1/2}=\Phi M^{1/2}U$, where $U\in C^{r\times N}$ is
a Right-unitary matrix, with $N$ and $r$ being the column number of $\Gamma $
and the rank of $\rho $. Therefore, based on the matrix notation and eq.
(9), eq. (23) can be directly rewritten in a twice-doubled Hilbert space as%
\begin{eqnarray}
\tau (\rho ) =\inf_{U}\sum_{i=1}^{N}\left\{ [\left( U^{T}\otimes
U^{T}\otimes U^{\dag }\otimes U^{\dag }\right) \right. \times \left.
\mathcal{A}\left( U\otimes U\otimes U^{\ast }\otimes U^{\ast
}\right) ]_{ii,ii}^{ii,ii}\right\} ^{\frac{1}{4}},
\end{eqnarray}%
where
\begin{equation}
\mathcal{A}=\left( \mathbf{\varrho }^{1/2}\right) ^{T}\sum_{i\neq l,j\neq
k}A_{ijkl}\left( \mathbf{\varrho }^{1/2}\right)
\end{equation}%
with%
\begin{equation}
\mathbf{\varrho }^{1/2}=\left( \Phi M^{1/2}\right) ^{T}\otimes \left( \Phi
M^{1/2}\right) ^{T}\otimes \left( \Phi M^{1/2}\right) ^{\dag }\otimes \left(
\Phi M^{1/2}\right) ^{\dag },
\end{equation}

\begin{equation}
\begin{array}{c}
A_{ijkl}=\Sigma _{ij}\otimes \Sigma _{lk}\otimes \Sigma _{ij}\otimes \Sigma
_{lk}+\Sigma _{ik}\otimes \Sigma _{lj}\otimes \Sigma _{ik}\otimes \Sigma
_{lj} \\
-\Sigma _{ij}\otimes \Sigma _{lk}\otimes \Sigma _{ik}\otimes \Sigma
_{lj}-\Sigma _{ik}\otimes \Sigma _{lj}\otimes \Sigma _{ij}\otimes \Sigma
_{lk},%
\end{array}%
\end{equation}%
and%
\begin{equation}
\Sigma _{ij}=\sigma _{y}\otimes \sigma _{y}\otimes e_{ij}.
\end{equation}%
If $\rho $ is defined in $C_{d\times d}$, $\mathcal{A}$ is then defined in $%
C_{d\times d}\otimes C_{d\times d}\otimes C_{d\times d}\otimes C_{d\times d}$%
. If the former two subspaces and the latter two ones are regarded
as a doubled subspace, respectively. $\mathcal{A}$ can be considered
to be defined in $C_{d^{2}\times d^{2}}\otimes C_{d^{2}\times
d^{2}}$. It is easy to find that $\mathcal{A}$ is invariant under
the exchange of two doubled subspaces. It is also obvious that
$\mathcal{A}$ is invariant, if the former two subspaces and the
latter two ones are exchanged simultaneously. Due to the symmetry,
following the analogous procedure to that in Ref. [16], we have the
following relations by means of kronecker approximation technique
[16,26,27].
\begin{equation}
\mathcal{A=}\sum_{i=1}^{r^{\prime }}B_{i}\otimes B_{i}
\end{equation}%
with $B_{i}$ defined in $C_{d\times d}\otimes C_{d\times d}$, and
\begin{equation}
B_{i}=\sum_{j=1}^{r^{\prime \prime }}\left( \sigma _{i}\right)
_{j}\left( C_{i}\right) _{j}\otimes \left( C_{i}\right) _{j}
\end{equation}%
with $\left( C_{i}\right) _{j}$ defined in $C_{d\times d}$, and
$\left( \sigma _{i}\right) _{j}$ being the corresponding singular
values [28]. Substitute above relations into eq. (25), one can
obtain that
\begin{equation}
\tau (\rho )=\inf_{U}\sum_{i=1}^{N}\left( \sum_{j=1}^{r^{\prime
}}\left( \sum_{m}^{r^{\prime \prime }}\left\vert \left( U^{T}\left(
C_{j}\right) _{m}U\right) _{ii}\right\vert ^{2}\right) ^{2}\right)
^{1/4}.
\end{equation}%
Following the procedure of Ref. [16] again, one can also obtain
three lower bounds, which have the same form to those in Ref.
$[16]$. Therefore, we do not give these bounds here.

Similarly, one can also find that the numerical realization to calculate the
bounds for a mixed state $\rho $ faces the same problem mentioned in Ref. $%
[16]$, i.e. the lower efficiency of calculation. To avoid the
problem, again we employ the method given in Ref. [29] to present an
analytic approximation of eq. (32) for weakly mixed states-quasi
pure states. In this way, we can conveniently demonstrate the
applications of our measure to some quasi pure states.

Analogous to Ref. [29], the tensor $\mathcal{A}$ can be obtained by

\begin{eqnarray}
&&\mathcal{A}_{p^{\prime }m^{\prime },n^{\prime }q^{\prime
}}^{pm,nq}=\sum_{i\neq l,j\neq
k}\sqrt{u_{p}u_{m}u_{n}u_{q}u_{p^{\prime }}u_{m^{\prime
}}u_{n^{\prime }}u_{q^{\prime }}} \times \left[ \left\langle \gamma
_{p}^{\ast }\right\vert \sigma _{y}\otimes \sigma _{y}\otimes
e_{ij}\left\vert \gamma _{p^{\prime }}\right\rangle \right. \notag \\
&&\times \left\langle \gamma _{m}^{\ast }\right\vert \sigma
_{y}\otimes \sigma _{y}\otimes e_{lk}\left\vert \gamma _{m^{\prime
}}\right\rangle -\left\langle \gamma _{p}^{\ast }\right\vert \sigma
_{y}\otimes \sigma _{y}\otimes e_{ik}\left\vert \gamma _{p^{\prime
}}\right\rangle \times \left. \left\langle \gamma _{m}^{\ast
}\right\vert \sigma _{y}\otimes \sigma _{y}\otimes e_{lj}\left\vert
\gamma _{m^{\prime }}\right\rangle \right]
  \notag\\
&&\times \left[ \left\langle \gamma _{n}\right\vert \sigma
_{y}\otimes \sigma _{y}\otimes e_{ij}\left\vert \gamma _{n^{\prime
}}^{\ast }\right\rangle \right. \times \left\langle \gamma
_{q}\right\vert \sigma _{y}\otimes \sigma _{y}\otimes
e_{lk}\left\vert \gamma _{q^{\prime }}^{\ast
}\right\rangle   \notag \\
&&-\left\langle \gamma _{n}\right\vert \sigma _{y}\otimes \sigma
_{y}\otimes e_{ik}\left\vert \gamma _{n^{\prime }}^{\ast
}\right\rangle \times \left. \left\langle \gamma _{q}\right\vert
\sigma _{y}\otimes \sigma _{y}\otimes e_{lj}\left\vert \gamma
_{q^{\prime }}^{\ast }\right\rangle \right] ,
\end{eqnarray}%
where $\left\vert \gamma _{\alpha }\right\rangle$ and $u_{\alpha }$
denote the $\alpha $th eigenvector and eigenvalue of $\rho $
respectively and all the other quantities are defined similar to
those in eq. (9). According to the symmetry of $A$ and the kronecker
product approximation technique in above section, $A$ can be
formally written as
\begin{equation}
\mathcal{A}_{p^{\prime }m^{\prime },n^{\prime }q^{\prime
}}^{pm,nq}=\sum_{\alpha }T_{pm}^{\alpha }\left( T_{p^{\prime }m^{\prime
}}^{\alpha }\right) ^{\ast }T_{nq}^{\alpha }\left( T_{n^{\prime }q^{\prime
}}^{\alpha }\right) ^{\ast }.
\end{equation}%
The density matrix of quasi pure states has one single eigenvalue $\mu _{1}$
that is much larger than all the others, which induces a natural order in
terms of the small eigenvalues $\mu _{i}$, $i>1$. Due to the same reasons to
those in Ref. [29], here we consider the second order elements of type $%
A_{11,11}^{pm,11}$. Therefore, one can have the approximation
\begin{equation}
\mathcal{A}_{p^{\prime }m^{\prime },n^{\prime }q^{\prime }}^{pm,nq}\simeq
\kappa _{pm}\kappa _{p^{\prime }m^{\prime }}\kappa _{nq}^{\ast }\kappa
_{n^{\prime }q^{\prime }}^{\ast }\ with \ \kappa _{pm}=\frac{\mathcal{A}%
_{11,11}^{pm,11}}{\sqrt[4]{\left( \mathcal{A}_{11,11}^{11,11}\right) ^{3}}}.
\end{equation}%
In this sense, eq. (32) can be simplified significantly:%
\begin{equation}
\tau (\rho )\simeq \tau _{a}(\rho )=\inf_{U}\sum_{i}\left\vert U^{T}\kappa
U\right\vert _{ii}.
\end{equation}%
$\tau _{a}(\rho )$ can be given by
\begin{equation}
\tau _{a}(\rho )=\max \{\lambda _{1}-\sum_{i>1}\lambda _{i},0\},
\end{equation}%
where $\lambda _{i}$ is the singular value of $\kappa $ in decreasing order.

As applications, let us consider two $\left( 2\times 2\times 3\right) -$%
dimensional quasi pure states constructed respectively by%
\begin{equation}
\rho _{1}(x)=x\left\vert GHZ^{\prime }\right\rangle \left\langle GHZ^{\prime
}\right\vert +(1-x)\mathbf{1}_{12}
\end{equation}%
and
\begin{equation}
\rho _{2}(x)=x\left\vert W^{\prime }\right\rangle \left\langle W^{\prime
}\right\vert +(1-x)\mathbf{1}_{12},
\end{equation}%
where
\begin{equation}
\left\vert GHZ^{\prime }\right\rangle =\frac{1}{2}(\left\vert
000\right\rangle +\left\vert 101\right\rangle +\left\vert 011\right\rangle
+\left\vert 112\right\rangle ),
\end{equation}%
and
\begin{equation}
\left\vert W^{\prime }\right\rangle =\frac{1}{\sqrt{3}}\left( \left\vert
000\right\rangle +\left\vert 011\right\rangle +\left\vert 112\right\rangle
\right) .
\end{equation}%
Note that $\left\vert GHZ^{\prime }\right\rangle $ and $\left\vert W^{\prime
}\right\rangle $ given in Ref. [11] correspond to $GHZ$ class and $W$ class
with high local rank, respectively. The two states can be considered as
quasi pure states for $x\geq 0.3$. $\tau _{a}$ for $\rho _{1}(x)$ and $\rho
_{2}(x)$ are both shown in Fig. 1, where the solid line corresponds to $\tau
_{a}(\rho _{1})$ and the dotted line corresponds to $\tau _{a}(\rho _{2})$.
Fig. 1 shows the sufficiency to test genuine tripartite entanglement for
such quasi pure states. In this sense, the measure presented in the paper
can characterize the properties of genuine tripartite entanglement and can
serve as an effective indicator of genuine tripartite entanglement.

\begin{figure} [htbp]
\centerline{\epsfig{file=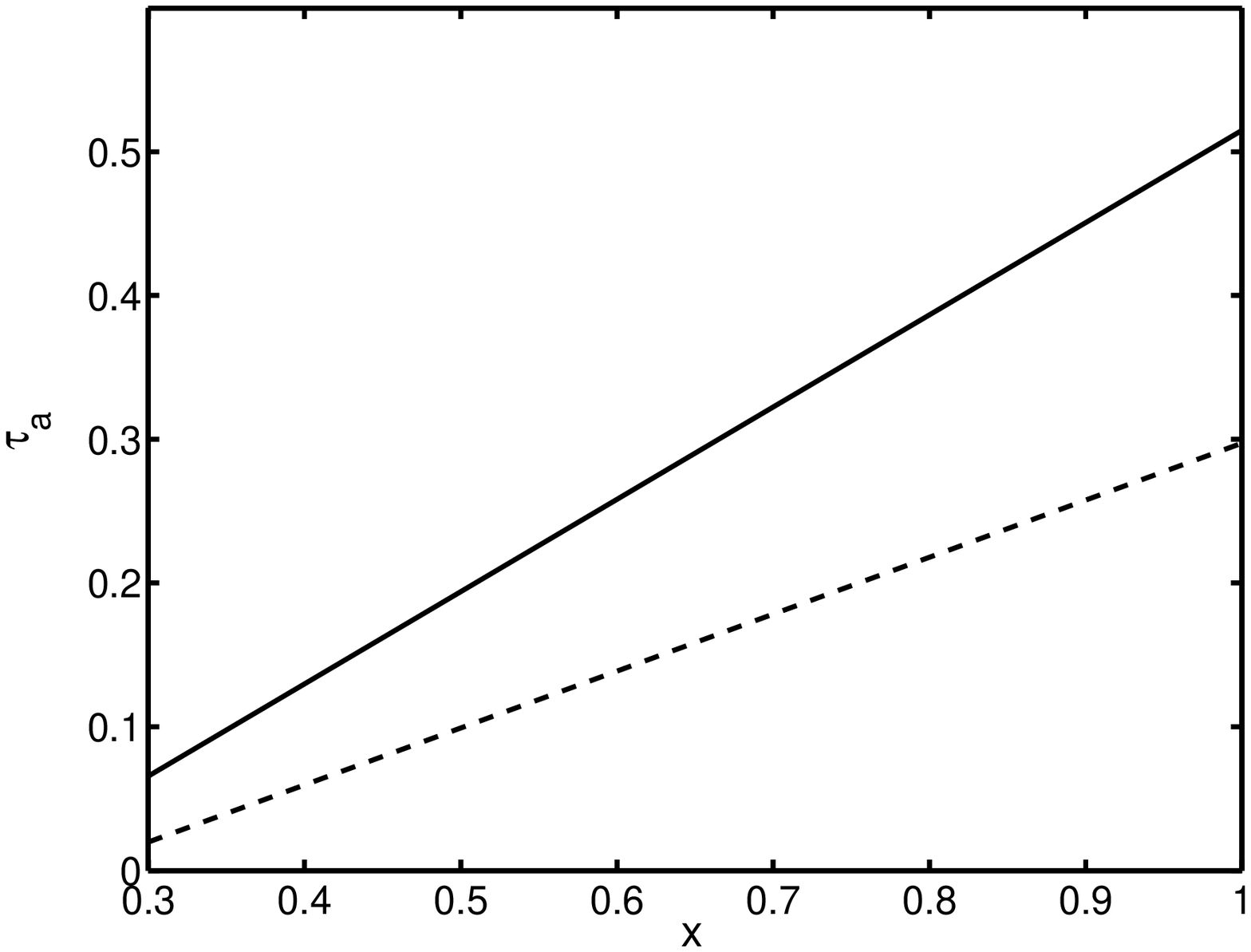, width=8.5cm}} 
\vspace*{13pt} \fcaption{\label{motion}$\protect\tau _{a}$ (dimensionless) for quasi pure states $\protect%
\rho _{1}(x)=x\left\vert GHZ^{\prime }\right\rangle \left\langle
GHZ^{\prime}\right\vert +(1-x)\mathbf{1}_{12}$ (solid line) and $\protect\rho %
_{2}(x)=x\left\vert W^{\prime }\right\rangle \left\langle W^{\prime
}\right\vert +(1-x)\mathbf{1}_{12}$ (dotted line) $vs$ $x$, $x\in
\lbrack 0.3,1]$.}
\end{figure}

\section{Conclusion and Discussion}

In summary, we have introduced an entanglement semi-monotone $\tau $
by a new approach to measure the genuine tripartite entanglement
existing in a given tripartite $(2\times 2\times n)-$dimensional quantum pure states. For $%
(2\times 2\times 2)-$dimensional systems, $\tau $ can be reduced to
the initial residual entanglement given in Ref. [9], but there
exists a neglectable constant difference between them. In
particular, it does not vanish for $W$ states with high local rank.
In this sense, $\tau $ can be considered as a generalization of the
initial residual entanglement. What's more, $\tau $ can conveniently
extended to the case of mixed states by utilizing the kronecker
product approximation technique. For the weakly mixed states, i.e.
quasi pure states, we have provided an analytic approximation, by
which we have investigated the genuine tripartite entanglement of
two quasi pure states. Numerical results show that $\tau$ obtained
from the analytic approximation can serve as an effective indicator
of genuine tripartite entanglement for quasi pure states. Our result
can be generalized to $(d\times d\times n)-$dimensional systems, but
many details are quite different and the property of entanglement
monotone might be lost. We would like to study the generalization in
detail elsewhere. What's more, it will be more valuable that $\tau$
for pure states can be employed to signal the phase transition of
some spin interaction systems by considering tripartite
entanglement, which is our forthcoming work.

\section{Acknowledgement}

This work was supported by the National Natural Science Foundation of China,
under Grant Nos. 10575017 and 60472017.


\begin{thebibliography}{000}
\bibitem{[1]} M. A. Nielsen and I. L. Chuang, \textit{Quantum Computation
and Quantum Information} (Cambridge University Press, Cambridge, 2000).

\bibitem{[2]} M. Zukowski, A. Zeilinger, M. A. Horne, and A. K. Ekert, Phys.
Rev. Lett. \textbf{71}, 4287 (1993).

\bibitem{[3]} C. H. Bennett, et al., Phys. Rev. Lett.\textbf{70},1895 (1993).

\bibitem{[4]} C. H. Bennett and S. Wiesner, Phys. Rev. Lett. \textbf{69},
2881 (1992).

\bibitem{[5]} W. K. Wootters, Phys. Rev. Lett. \textbf{80}, 2245 (1998).

\bibitem{[6]} A.Uhlmann, Phys. Rev. A \textbf{62}, 032307 (2000).

\bibitem{[7]} K. Audenaert, F. Verstraete and De Moor, Phys. Rev. A \textbf{64%
}, 052304 (2001).

\bibitem{[8]} Florian Mintert, Marek Ku\'{s}, and Andreas Buchleitner, Phys.
Rev. Lett. \textbf{92}, 167902 (2004).

\bibitem{[9]} Valerie Coffman, Joydip Kundu, and William K. Wootters, Phys.
Rev. A \textbf{61}, 052306 (2000).

\bibitem{[10]} W. D\"{u}r, G. Vidal, and J. I. Cirac, Phys. Rev. A \textbf{62%
}, 062314 (2000).

\bibitem{[11]} A. Miyake, Phys. Rev. A \textbf{67}, 012108 (2003).

\bibitem{[12]} A. Miyake, F. Verstraete, Phys. Rev. A \textbf{69},
012101 (2004).

\bibitem{[13]} Andreas Osterloh, Jens Siewert, Phys. Rev. A \textbf{72},
012337 (2005).

\bibitem{[14]} Alexander Wong and Nelson Christensen, Phys. Rev. A \textbf{63%
}, 044301 (2001).

\bibitem{[15]} Chang-shui Yu, He-shan Song, Phys. Rev. A \textbf{72}, 022333
(2005).

\bibitem{[16]} Chang-shui Yu, He-shan Song, Phys. Rev. A \textbf{73},
032322 (2006).

\bibitem{[17]} A. R. R. Carvalho, F. Mintert, A. Buchleitner, Phys. Rev.
Lett. \textbf{93,} 230501 (2004);

\bibitem{[18]} Chang-shui Yu, He-shan Song, Phys. Rev. A \textbf{73}, 022325
(2006).

\bibitem{[19]} D. A. Meyer and N. R. Wallach, J. Math. Phys. \textbf{43},
4273 (2002).

\bibitem{[20]} G. K. Brennen, Quant. Inf. Comp. \textbf{3}, 619 (2003).

\bibitem{[21]} G. Gour, D. A. Meyer, and B. C. Sanders, Phys. Rev. A \textbf{72},
042329 (2005).

\bibitem{[22]} G. Gour, Phys. Rev. A \textbf{72}, 042318 (2005).

\bibitem{[23]} P. Rungta, V. Bu\v{z}ek, C. M. Caves, et al, Phy. Rev. A%
\textbf{\ 64}, 042315 (2001).

\bibitem{[24]} W. K. Wootters, Quantum Inf. Comp. \textbf{1}, 27 (2001).

\bibitem{[25]} In fact, the entanglement of $W$ states is a kind of genuine tripartite
entanglement [10-12], however, $\tau (W)=0$ for $\left( 2\times
2\times 2\right) -$dimensional cases. In this sense, the genuine
tripartite entanglement mentioned in the paper means those with
$nonzero$ $\tau $.

\bibitem{[26]} N. P. Pitsianis, Ph.D. thesis, Cornell University, New York,
1997.

\bibitem{[27]} C. F. Van Loan and N. P. Pitsianis, in \textit{Linear Algebra
for Large Scale and Real Time Applications}, edited by M. S. Moonen and G.
H. Golub (Kluwer, Dordrecht, 1993), pp. 293-314.

\bibitem{[28]} The procedure is the same to that in Ref. [15], hence omitted
here. All corresponding parameters are not given explicitly, because they
are usually given by numerical processing.

\bibitem{[29]} Florian Mintert, Andr\'{e} R. R. Carvalho, Marek Ku\'{s}, and
Andreas Buchleitner, Physics Report \textbf{415}, 207 (2005).
\end{thebibliography}
\end{document}